\documentclass[intlimits,twoside,a4paper]{article}

\usepackage[T2A,T1]{fontenc} %%%% already in cmpj3.sty

\usepackage[cp1251]{inputenc}
\usepackage[eqsecnum]{cmpj3}

\usepackage{bm} 
\usepackage{ulem}
\usepackage{xcolor}

\articletype{A part of the Special collection to the 100th anniversary of the birth of Ihor Yukhnovskii}

\issue{2025}{28}{3}{33502}
\doinumber{10.5488/CMP.28.33502}
\title[Thermodynamic relation for the systems with inhomogeneous distribution of particles]%
{Thermodynamic relation for the systems with inhomogeneous distribution of particles}
\author[A. P. Rebesh, B. I. Lev, A. G. Zagorodny]{A. P. Rebesh\orcid{0000-0002-6784-1889}\thanks{\email{Corresponding author: nrebesh@gmail.com}.},
        B. I. Lev\orcid{0000-0003-3905-2070}, A. G. Zagorodny\orcid{0000-0002-7953-6726}}
\address{Bogolyubov Institute for Theoretical Physics of the NAS of Ukraine, 14-b Metrolohichna Str., 
03143 Kyiv, Ukraine}


\date{Received March 13, 2025, in final form June 18, 2025}

\begin{document}

\maketitle

\begin{abstract}
For the system with inhomogeneous distribution of macroscopic parameters we obtain thermodynamic relation which depends on the spatial point (coordinate). In our approach, to obtain such a relation we use the basic ideas of the method of nonequilibrium statistical operator combined with the Hubbard--Stratonovich transformation. First of all, we define the thermodynamic relation for the system with homogeneous distribution of particles. Possible behavior peculiarities of systems with different character of interaction in nonequilibrium case are predicted. By saddle-point method we find the dominant contributions to the partition function and obtain all thermodynamic parameters of the systems with different character of interaction. The formations of saddle state in all systems of interacting particles at different temperatures and particle distributions have the same physical nature and therefore they can be described in the same way. We consider the systems with attractive and repulsive interactions as well as self-gravitating systems. 
%
%
%\keywords Local equilibrium partition function, equation of state, self-gravitating system, long-range interaction
\printkeywords
%
%\pacs Up to six PACS numbers (optional)
\end{abstract}

\section{Introduction}
This article is dedicated to the 100th anniversary of the birth of the outstanding Ukrainian theoretical physicist, teacher, public and state figure Ihor Yukhnovskii. He made a significant contribution to the development of statistical theory and thermodynamics of condensed matter systems, using and developing various methods of theoretical study of such systems --- the method of collective variables, the method of displacement of collective variables, the method of renormalization group, the method of cluster expansions, etc. (see, for example, \cite{IR-1,IR-2,IR-3,IR-4}). One of such methods is the method of nonequilibrium statistical operator, proposed by D. Zubarev \cite{Zub}. Using the combination of this method with the kinetic approach, I.~Yukhnovskii proposed a consistent description of nonlinear hydrodynamic fluctuations in many-particle systems \cite{IR-5}. The basic principles of the method of the nonequilibrium statistical operator is used in our article in order to describe quasi-equilibrium inhomogeneous steady states.

Statistical description of interacting particles has attracted a permanent attention. The study of interacting particle systems promotes the development of fundamental ideas of statistical mechanics and thermodynamics and enables their testing. A few model systems of interacting particles are known, as far as the partition function can be exactly evaluated, at least, in the thermodynamic limit. Particle systems with long-range interaction sometimes cannot be described in terms of a usual thermodynamic ensemble~\cite{Tir,Cav}. It means that such systems cannot be treated by standard methods of equilibrium statistical mechanics. In particular, inasmuch as energy is non-additive, the canonical ensemble is inapplicable of  studying  the systems with long-range interactions. The equilibrium states are only local thermodynamic potential maximum \cite{Zub}. The system is stable but thermodynamic limit does not exist \cite{Lal}. In order to determine the equilibrium states of the systems and to describe the probable phase transitions, the two approaches, statistical and thermodynamic, have been developed. It is generally believed that for such systems the mean-field theory is exact. In this approach any thermodynamic function depends on the parameters only in terms of dimensionless combinations. 

The systems with long-range interactions, e.g., self-gravitating systems or systems with Coulomb repulsive interaction, do not relax towards usual Boltzmann--Gibbs thermodynamic equilibrium. They get trapped in quasi-stationary states whose lifetimes diverge as the number of particles increases. A quantitative description of the instability threshold of spontaneous symmetry breaking for $d$-dimensional systems is given in reference \cite{Pak}. The homogeneous particle distribution in interacting systems can be unstable (for example, in the case of systems with long-range attraction potentials) and spatial inhomogeneity can appear from the beginning. The description of interacting systems within various equilibrium ensembles should be done in different ways, especially if the quasi-stationary state of the system is far from equilibrium and the time of relaxation towards equilibrium state is very long. Nonequilibrium stationary states were described in reference \cite{Ben}. It was shown that three-dimensional systems is trapped in quasi-stationary states rather than evolve towards thermodynamic equilibrium. 

Thus, a dilemma arises, either to employ the postulates of equilibrium statistical mechanics and obtain only instability criteria, or to use different approaches which permit  the space distributions to be inhomogeneous. The particle density, temperature and chemical potential in the case of macroscopically inhomogeneous systems can be described in terms of the nonequilibrium statistical operator approach~\cite{Zub} that takes into account the probability of local changes of the thermodynamic parameters. Such an approach does not seem to be consistent inasmuch as the state equation should follow from the partition function which is unknown \cite{Bax,Rue}. The system is nonequilibrium a priori and inhomogeneity of the distribution of particles can induce distributions of temperature, chemical potential and other thermodynamic parameters. Using the nonequilibrium statistical operator approach together with the Hubbard--Stratonovich transformation \cite{Hubbard,Str} makes it possible to describe such systems.

The formation of a spatially inhomogeneous distribution of interacting particles is a typical problem in condensed matter physics. It requires a non-conventional statistical description of a system of interacting particles to be tailored in a way to allow for an arbitrary spatially inhomogeneous  distribution of particles. The statistical description should employ a procedure of calculation of the dominant contributions to the partition function and to avoid the local thermodynamic potential divergences for an infinite system volume. A non-conventional method for the treatment of the problem was proposed in references \cite{Bel,Lev,Lev1,Kle}. It employs the Hubbard--Stratonovich representation of the partition function \cite{Hubbard,Str}, that is extended and applied to the interacting particles systems to find the solution for the  distribution of particles without necessity to use the spatial box restrictions. It is important that this solution has no divergences in the thermodynamic limits. For this purpose, one can use the saddle-point approximation with regard to the conservation of the number of particles in the limited space. The partition functions for both cases of homogeneous and inhomogeneous  distributions of particles were obtained in \cite{Lev,Lev1}. This approach, however, provides only the condition for the formation of probable inhomogeneous distributions in a system of interacting particles. 

In this article we develop a new approach to considering the systems of interacting particles in local equilibrium states based on the method of the nonequilibrium statistical operator proposed by Zubarev~\cite{Zub} combined with the Hubbard--Stratonovich transformation \cite{Hubbard,Str}. The particular interest of our approach may consist in the possibility to obtain the necessary relations which are correct for the case of homogeneous particle distribution (which is shown in the present article) as well as for the case of inhomogeneous particle distribution. What is important is that our approach permits to obtain thermodynamic functions when temperature and chemical potential depend on the spacial point (coordinate), while traditional thermodynamics does not assume spatial dependence of thermodynamic relations.    
The method of the Zubarev statistical operator itself has been widely used for the description of physical systems. As one of examples, it was used to describe the generalized hydrodynamic state of a magnetic fluid in an external magnetic field \cite{Mryglod}. 

The equation of state and all necessary thermodynamic characteristics are governed by the equations which contribute mainly to the partition function. Therefore, there is no need to introduce an additional hypothesis about the density-dependence of the temperature. The last dependence follows from the solution of the relevant thermodynamic relations that yield the extreme values of the nonequilibrium partition function. Such an approach was proposed earlier to describe a spatially inhomogeneous distribution in self-gravitating systems where the only attractive interaction is considered \cite{Lev1,Lev2,Lev4,Lev3}.
The long-term stellar dynamical evolution of self-gravitating systems on the time scales, that are  much longer than the two-body relaxation time, was also studied \cite{Taruya}. 
However, in our present article we 
describe the system with simultaneous attractive and repulsive interaction in the framework of our proposed approach. We also derive  the general local thermodynamic relations for the macroscopic parameters of the system and determine the conditions for realization of spatially inhomogeneous states on the thermodynamic limit. 
The relations which we get in the framework of the local equilibrium partition function approach can be used for  obtaining  the local equations of state. In general, the physical applications of local equilibrium partition function 
 are well described  in references \cite{Zub,Zub1,Zub2}. 
 For the equilibrium conditions, the well-known result \cite{Hua,Isi} for the partition function is reproduced. We provide a detailed description of an interacting system (in general case with the different type of interaction between particles) based on the principles of nonequilibrium statistical mechanics and obtain the probable  distributions of particles for a fixed number of particles and energy of the systems. 
As a few examples of the systems with inhomogeneous  distribution of particle, which can be described in the approach of local equilibrium partition function, we could mention the solid solutions, Martensitic alloys with the formation of corresponding structures \cite{Khach}, first order phase transitions with the formation of origins of a new phase \cite{Langer}. 
 
The paper is organized as follows. The general ideas of the proposed approach, based on the results of the local equilibrium partition function and the Hubbard--Stratonovich transformation, are presented in section~\ref{sec-2}. In section~\ref{sec-3}, we derive thermodynamic relations for spatially inhomogeneous systems. The application of the above approach to the description of long-range interacting systems (including the case of a self-gravitating system) is presented in section~\ref{sec-4}. The obtained results are briefly described in the Conclusions. The specific mathematical transformations are presented in the Appendix.

\section{Statistical description of nonequilibrium system of interacting particles}
\label{sec-2}

Statistical mechanics of nonequilibrium systems is based on the conservation laws  not for the average values of dynamic variables but  particularly for the dynamic variables. To determine the thermodynamic function of nonequilibrium systems, we should use the presentation of corresponding statistical ensembles which take into account the nonequilibrium states of the systems. In such a case,  one can determine the nonequilibrium ensemble 
as the totality of a system that contains the same stationary external action. This 
system has the same character of contact with thermostat and possesses all possible values of macroscopic parameters. In the system, which is in the same stationary external condition, the local equilibrium stationary distribution will be formed. If the external condition depends on time, then the local equilibrium distribution is not stationary. The local equilibrium ensemble must accordingly determine the distribution function or statistical operator of the system \cite{Zub}. Finally, the stable states on the series of equilibrium  classical particles are only metastable because they correspond to the local maximum of the local thermodynamic potential from which the behavior of the systems can be determined.

For the classical case, the local distribution function can be written as \cite{Zub,Tok}:
\begin{equation}  \label{eq1} 
P_N(x^N,t)=Q^{-1}_l\exp\left\{-\int \left[\beta(\mathbf{r})H(\mathbf{r})-\eta(\mathbf{r})n(\mathbf{r})\right]\rd\mathbf{r}\right\},
\end{equation}
where $Q_l$ is the statistical sum of the local equilibrium distribution:
\begin{equation}      \nonumber
Q_{l}=\int D\Gamma \exp\left\{-\int
\left[\beta(\mathbf{r})H(\mathbf{r})-\eta(\mathbf{r})n(\mathbf{r})\right]\rd\mathbf{r}\right\},
\end{equation} 
where the inverse temperature $\beta$ and chemical potential $\eta$ depend on the spatial point and $H(\mathbf{r})$ is the energy density.
The integration in the present formula (\ref{eq1}) must be performed over all phase space of the system.
We should note that in the case of the local equilibrium distribution, Lagrange multipliers 
$\beta(\mathbf{r})$ and $\eta(\mathbf{r})$ are the functions of the spatial point $\mathbf{r}$. The 
microscopic density of particles can be written in the standard form:
\begin{equation}      \label{eq2}
n(\mathbf{r})=\sum_{i}\delta(\mathbf{r}-\mathbf{r_{i}}).
\end{equation} 
It is possible to introduce the local equilibrium distribution if the relaxation time 
in all systems is larger than the relaxation time of the local macroscopical area which is a part of 
this system. The conservation number of particles and energy in the systems can be presented in the form of usual relations: $\int n(\mathbf{r})\,\rd\mathbf{r} = N$ and $ \int  H(\mathbf{r})\,\rd \mathbf{r} = E$. When the local equilibrium partition function is determined, one can obtain all thermodynamic parameters of the nonequilibrium systems. Phenomenological ther\-mo\-dy\-na\-mics is based on the conservation laws for the average values of 
physical values, such as the number of particles, energy and momentum. To this end, one can determine the thermodynamic relation for inhomogeneous systems. After variation of the statistical sum by Lagrange multipliers, we can write the necessary thermodynamic relations in the form \cite{Zub}:
\begin{equation}   \label{eq3}
-\frac{\delta \ln Q_{l}}{\delta \beta(\mathbf{r})}=\langle
H(\mathbf{r})\rangle_{l}, \hspace{5mm} \frac{\delta \ln Q_{l}}{\delta \eta (\mathbf{r})}=\left\langle
n(\mathbf{r})\right\rangle_{l}.
\end{equation} 
The relations in (\ref{eq3}) are a general extension of the well-known relation for the equilibrium systems to the case of inhomogeneous systems.

For further studies, it is natural to determine the quantity
\begin{equation}     \label{eq4}
H(\mathbf{r})=\sum_{i}\bigg[\frac{\mathbf{p}_{i}^{2}}{2m}-\frac{1}{2}\sum_{j} W(\mathbf{r}_i,\mathbf{r}_j)+\frac{1}{2}\sum_{j}
U(\mathbf{r}_i,\mathbf{r}_j)\bigg]\delta(\mathbf{r}-\mathbf{r}_i) ,
\end{equation}
as a dynamic variable of the energy density. Then, obviously, the Hamiltonian of the system $H$ will be as follows:
\begin{equation} 
H=\int H(\mathbf{r})\,\rd\mathbf{r},
\end{equation}
with attractive $W(\mathbf{r_{i}},\mathbf{r_{j}})$ and repulsive $U(\mathbf{r_{i}},\mathbf{r_{j}})$ parts of the interaction energy, correspondingly (see also Appendix). 

Using the Hubbard--Stratonovich transformation, now we rewrite the statistical sum of the local equilibrium distribution in terms of the additional fields as
\begin{equation}   \label{eq7}
Q_{l}=\int D\varphi D\psi \,\rd \xi
\exp \left[-S(\varphi(\mathbf{r}),\psi(\mathbf{r}),\xi(\mathbf{r}),\beta(\mathbf{r}))\right],
\end{equation}
where the effective local thermodynamic potential can be written in the following form:
\begin{align}     \label{eq8} 
S &=\frac{1}{2}\iint 
U(\mathbf{r},\mathbf{r'})^{-1}\varphi(\mathbf{r})\varphi(\mathbf{r'})\,\rd\mathbf{r}\,\rd \mathbf{r'}-\frac{1}{2}\iint W(\mathbf{r},\mathbf{r'})^{-1}\psi(\mathbf{r})\psi(\mathbf{r'})\,\rd\mathbf{r}\,\rd \mathbf{r'} \nonumber \\ &
-\int \bigg\{
\xi(\mathbf{r})\left[\frac{2\piup m
	%(\mathbf{r})
}{\hbar^{3}\beta(\mathbf{r})}
\right]^{{3}/{2}}\exp\left(\sqrt{\beta(\mathbf{r})}\psi(\mathbf{r})\right)\cos \left[\sqrt{\beta(\mathbf{r})}\varphi(\mathbf{r})\right]\bigg\}\,\rd\mathbf{r}.
\end{align}
The mathematical manipulations which were done to derive the expression (\ref{eq7}) are presented in Appendix.
The functional $S\equiv S(\varphi(\mathbf{r}), \psi(\mathbf{r}), \xi(\mathbf{r}),\beta(\mathbf{r}))$ depends on the distribution of the field variables 
$\varphi(\mathbf{r})$ and $\psi(\mathbf{r})$ which describe the repulsive and attractive interaction, the chemical activity $\xi(\mathbf{r})$ and inverse temperature $\beta(\mathbf{r})$.  
To find the asymptotic value of the statistical sum $Q_{l}$ for an increasing number of particles $N\to\infty$ we  use the saddle-point method. 
It was shown that the solution in saddle-point can be viewed as the mean field approximation \cite{Bel,Lev}.
Representation of the partition function in terms of the functional integral over the auxiliary fields corresponds to the construction of an equilibrium sequence of the probable states with regard to their weights. In such a case, the methods of quantum field theory can be employed. The extension to a complex plane provides a possibility to apply the saddle-point method and to select the system states whose contributions in the partition function are dominant \cite{Lev}.

The dominant contribution is given by the states which satisfy the extreme condition for the functional. It is easy to see that the saddle-point equation presents the thermodynamic relation, and it can be written in the form of equation for the field variables: 
\begin{equation}    \label{eq9} 
\frac{\delta S}{\delta \varphi(\mathbf{r})}=0, \hspace{5mm} \frac{\delta S}{\delta \psi(\mathbf{r})}=0, 
\end{equation}
for the normalization condition:
\begin{equation}   \label{eq10} 
\frac{\delta S}{\delta \eta(\mathbf{r})}=-\int_V \frac{\delta
	S}{\delta \xi(\mathbf{r})}\xi(\mathbf{r})\,\rd\mathbf{r}=N,
\end{equation}
and for the conservation of the energy of the system:
\begin{equation}  \label{eq11} 
-\int \frac{\delta S}{\delta \beta (\mathbf{r})}\,\rd\mathbf{r}=E.
\end{equation}
The solution of the above equation (\ref{eq9}) fully determines all macroscopic thermodynamic parameters and describes the general behavior of  the system of interacting particles. It does not matter whether the distribution of particles is spatially inhomogeneous or homogeneous. The above set of equations~(\ref{eq9})--(\ref{eq11}) in principle solves the many-particle problem in the thermodynamic limit. The spatially inhomogeneous solution of these equations corresponds to the distribution of the interacting particles. Such an inhomogeneous behavior is associated with the nature and intensity of the interaction. In other words, accumulation of the particles in a finite spatial region (formation of a cluster) reflects the spatial distribution of the fields, the activity and temperature. It is a very important note, that only in this approach we can take into account the inhomogeneous distribution of the temperature and chemical potential, which can depend on the spatial distribution of particles in the system. 

\section{Thermodynamic relation}
\label{sec-3}

For further consideration we  analyze the general form of the local thermodynamic potential~\eqref{eq8}. We introduce the new field variables $\Phi(\mathbf{r})=\sqrt{\beta(\mathbf{r})}\varphi(\mathbf{r})$, $\Psi(\mathbf{r})=\sqrt{\beta(\mathbf{r})}\psi(\mathbf{r})$ and rewrite the local thermodynamic potential in a more simple form:  
\begin{align}    \label{eq12}    
S =\,&\frac{1}{2}\iint 
[\beta(\mathbf{r})U(\mathbf{r},\mathbf{r'})]^{-1}\Phi(\mathbf{r})\Phi(\mathbf{r'})\, \rd\mathbf{r}\, \rd \mathbf{r'}  -   \frac{1}{2}\iint[\beta(\mathbf{r}) W(\mathbf{r},\mathbf{r'})]^{-1}\Psi(\mathbf{r})\Psi(\mathbf{r'})\, \rd\mathbf{r}\, \rd \mathbf{r'}    \nonumber \\ & -
\int \left[
\xi(\mathbf{r})\Lambda^{-3}(\mathbf{r}) \exp \Psi(\mathbf{r})\cos \Phi(\mathbf{r})\right]\rd\mathbf{r},  
\end{align} 
where $\Lambda(\mathbf{r})=\left[{\hbar^{2}\beta(\mathbf{r})}/{2 m} \right]^{{1}/{2}} $ 
is the local thermal de~Broglie wavelength. From normalization condition~(\ref{eq10}) we can obtain the relation
\begin{equation}   \label{eq13}
\int \left[
\xi(\mathbf{r})\Lambda^{-3}(\mathbf{r}) \exp \Psi(\mathbf{r})\cos \Phi(\mathbf{r})\right]\rd\mathbf{r}=N,
\end{equation}
which provides the macroscopic density function $\rho(\mathbf{r})$ given by
\begin{equation}      \label{eq14}
\rho(\mathbf{r})\equiv \Lambda^{-3}(\mathbf{r}) \xi(\mathbf{r})\cos\big[\sqrt{\beta(\mathbf{r})}\varphi(\mathbf{r})\big]\exp\big[\sqrt{\beta(\mathbf{r})\psi(\mathbf{r})}\big]. 
\end{equation}
In the case of the system without interaction (for free particles $ \Phi(\mathbf{r})=0 $, $\Psi(\mathbf{r})=0$), if we write the chemical activity in terms of the chemical potential as $\xi(\mathbf{r})=\exp [\mu(\mathbf{r})\beta(\mathbf{r})]$, we
can obtain the well-known relation: $\beta(\mathbf{r})\mu(\mathbf{r})=\ln \rho(\mathbf{r})\Lambda^{3}(\mathbf{r})$ which generalizes the relation for the equilibrium statistical mechanics \cite{Hua}.

From thermodynamic relations (\ref{eq9}), written in  terms of the new variables $\Phi(\mathbf{r})$ and $\Psi(\mathbf{r})$ instead of $\varphi(\mathbf{r})$ and $\psi(\mathbf{r})$, we can obtain the equations
\begin{equation}     \label{eq15}
\hspace{-16mm} \text{for the field $\Psi(\mathbf{r})$:} \hspace{5mm} \int[\beta(\mathbf{r}) W(\mathbf{r},\mathbf{r'})]^{-1}\Psi(\mathbf{r'})\,\rd \mathbf{r'}+\rho(\mathbf{r})=0, 
\end{equation}
\begin{equation}     \label{eq16}
\text{for the field $\Phi(\mathbf{r})$:} \hspace{5mm}  \int[\beta(\mathbf{r})U(\mathbf{r},\mathbf{r'})]^{-1}\Phi(\mathbf{r'})\,\rd \mathbf{r'}+\rho(\mathbf{r}) \tan [\Phi(\mathbf{r'})]=0. 
\end{equation}
To define the field variables, we should multiply the first equation (\ref{eq15}) by $\int W(\mathbf{r},\mathbf{r'})\,\rd \mathbf{r'}$ and the second equation (\ref{eq16}) by $\int U(\mathbf{r},\mathbf{r'})\,\rd \mathbf{r'}$. As a result, we get
\begin{equation}   \label{eq17}     
\Psi(\mathbf{r})+\int \beta(\mathbf{r}) W(\mathbf{r},\mathbf{r'})\rho(\mathbf{r'}) \,\rd \mathbf{r'}=0 ,
\end{equation}
\begin{equation}    \label{eq18}  
\Phi(\mathbf{r'})+\int \beta(\mathbf{r})U(\mathbf{r},\mathbf{r'})\rho(\mathbf{r'}) \tan[\Phi(\mathbf{r'})] \,\rd \mathbf{r'}=0 .
\end{equation}
Using (\ref{eq17}), (\ref{eq18}) we obtain the local thermodynamic potential $\beta F=-\ln Q_{l}=S$ as
\begin{align}       \label{eq19}
S & =\beta F=\frac{1}{2}\iint
\beta(\mathbf{r}) U(\mathbf{r},\mathbf{r'})\rho(\mathbf{r}) \tan[\Phi(\mathbf{r})]\rho(\mathbf{r'})\tan[\Phi(\mathbf{r'})]\,\rd\mathbf{r}\,\rd \mathbf{r'} \nonumber \\ & - \frac{1}{2}\iint \beta(\mathbf{r}) W(\mathbf{r},\mathbf{r'})\rho(\mathbf{r})\rho(\mathbf{r'}) \,\rd\mathbf{r} \,\rd \mathbf{r'}-\int \rho(\mathbf{r})\,\rd\mathbf{r}. 
\end{align}
The quantity $S(\varphi(\mathbf{r})$, $\psi(\mathbf{r})$, $\xi(\mathbf{r})$, $\beta(\mathbf{r}))$ fully determines the free energy on the saddle state of the field variables. In this sense, the free energy is defined through the macroscopic parameter which is the solution of saddle equations (\ref{eq9}) and (\ref{eq10}) and determines the thermodynamic stable states of the system. From local nonequilibrium free energy we can also obtain  the equation of state of the nonequilibrium system:
\begin{align}     \label{eq20}
P(\bf{r}) & =- \frac{\partial F}{\partial V}=\frac{1}{\beta(\mathbf{r})}\rho(\mathbf{r})+\int W(\mathbf{r},\mathbf{r'})\rho(\mathbf{r})\rho(\mathbf{r'})\,\rd \mathbf{r'} \nonumber \\ &- \int
U(\mathbf{r},\mathbf{r'})\rho(\mathbf{r}) \tan[\Phi(\mathbf{r})]\rho(\mathbf{r'})\tan[\Phi(\mathbf{r'})] \,\rd \mathbf{r'}.
\end{align}
Equation (\ref{eq20}) is the local equation of  state of nonequilibrium system. Without interaction in the system, we can reproduce the well-known equation of  state $P={\rho(\mathbf{r})}/{\beta(\mathbf{r})}$, from which it is easy to see that the pressure can be negative if the value of $\tan[\Phi(\mathbf{r})]$ increases. This becomes possible if the field which describes the repulsive interaction has a specific value which is equal to $\Phi=\frac{\piup}{2}-\delta $, where $\delta^2 \sim -\frac{2}{\piup}\int
\beta U(\mathbf{r},\mathbf{r'})\rho(\mathbf{r'}) \,\rd \mathbf{r'}$, and it is attained by the special character of repulsive interaction. From such determination we can conclude that the real local thermodynamic potential can be presented in the general form: 
\begin{equation}     \label{eq21}
\beta F=-\frac{1}{2}\int \left[P(\mathbf{r})+ \rho(\mathbf{r})\right] \,\rd\mathbf{r}=-\frac{1}{2}\bigg[N+\int P(\mathbf{r})\,\rd\mathbf{r}\bigg],
\end{equation}
which in the general case automatically determines the real local free energy of the system of particles $F=\frac{1}{2}Nk_{\rm B}T+ \int k_{\rm B}T P(\mathbf{r}) \,\rd\mathbf{r}$, where $k_{\rm B}$ is the Boltzmann constant.  
The chemical potential can be written as functional \cite{Lif,Gor}:
\begin{align}       \label{eq22}
\mu(\mathbf{r}) & =\frac{\delta F}{\delta N}= \frac{1}{V}\frac{\delta F}{\delta \rho(\mathbf{r})}=\frac{1}{\beta(\mathbf{r})}+\int W(\mathbf{r},\mathbf{r'})\rho(\mathbf{r'})\,\rd \mathbf{r'}  \nonumber \\ & -  \int
U(\mathbf{r},\mathbf{r'}) \tan[\Phi(\mathbf{r})]\rho(\mathbf{r'}) \tan[\Phi(\mathbf{r'})] \,\rd \mathbf{r'}.
\end{align}
Then, we have all the necessary thermodynamic relations for a description of the behavior of nonequilibrium system. The dynamic equation for evolution of the nonequilibrium system can be wrtten in the general form \cite{Cahn}: 
\begin{equation}         \label{eq23}
\frac{ \partial \rho(\mathbf{r})}{\partial t}=\nabla M(\mathbf{r}) \nabla \mu(\mathbf{r}),
\end{equation} 
where $M(\mathbf{r}) =D(\mathbf{r}) \beta(\mathbf{r})$ is Einstein's mobility which is determined through the diffusion coefficient $D(\mathbf{r})$. Such a  presentation leads to the determination of the evolution of nonequilibrium system in general case, which is impossible. Next, we deal only with the special case which makes it possible to obtain  particular results for few systems. Formally, we can introduce the critical temperature spinodal decomposition:
\begin{equation}    \label{eq24}
k_{\rm B}T_{\rm sd}=\int
U(\mathbf{r},\mathbf{r'}) \tan[\Phi(\mathbf{r})]\rho(\mathbf{r'})\tan[\Phi(\mathbf{r'})] \,\rd \mathbf{r'}-\int W(\mathbf{r},\mathbf{r'})\rho(\mathbf{r'})\,\rd \mathbf{r'},
\end{equation}
which is the motive evolution of the local density. If $T = T_{\rm sd}$ we have an equilibrium situation, and only for such a system any thermodynamic parameter does not change. The motive of any evolution of the systems is the difference of chemical potentials.

\begin{figure}[h]
	\begin{center}
		%\begin{minipage}[htb]{0.4\linewidth}
		\includegraphics[scale=0.7]{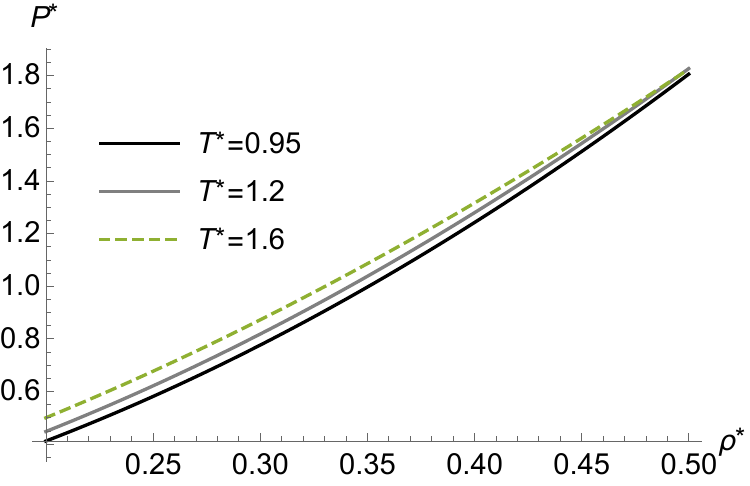} 
		\caption{{(Colour online) Dependence of $P^*$ on $\rho^*$ according to the equation of state (\ref{eq26}) for the reduced temperature of $T^*=0.95$ (113.8~K), $T^*=1.2$ (143.8~K), $T^*=1.6$ (191.7~K). Real units are for argon with $\sigma = 3.405$~\AA~and $\epsilon/k = 119.8$~K.}} \label{Fig1}
		%\end{minipage}
	\end{center}
\end{figure}

To conclude this section we consider, as a simple example, the case with constant temperature which is realized in the most conditions related to the equilibrium state. For constant density, we can write the free energy in a simple form:
\begin{equation}       \label{eq25}
 F=\frac{1}{2}V u \rho^2 \tan^{2}\Phi - \frac{1}{2}V  w \rho^2- \frac{1}{\beta}V\rho, 
\end{equation}
where $u=\int U(\mathbf{R})\,\rd\mathbf{R}$, $w=\int W(\mathbf{R})\,\rd\mathbf{R}$ and $\mathbf{R}=\mathbf{r}-\mathbf{r'}$ is radius-vector of the distance between the interacting particles. For field variables, we must solve the transcendent equation $\Phi+\beta u  \rho \tan\Phi=0$. As it is easy to see, in this special case we can write the equation of state in a very simple form:
\begin{equation}     \label{eq26}
P= \frac{1}{\beta}\rho+ \frac{1}{2}\rho^2 w -\frac{1}{2}\rho^2 u \tan^2\Phi.
\end{equation}
In figure~\ref{Fig1} for illustration we plot the equation of state (\ref{eq26}). To obtain the values of attractive and repulsive energy we take a Lennard-Jones potential:
\begin{equation} \label{26-1}
U  =  u(r)+w(r)=4\epsilon\left[  \left(\frac{\sigma}{r}\right)^{12}- \left(\frac{\sigma}{r}\right)^6 \right].
\end{equation}
Then, for the value of repulsive energy we have $u=\frac{16}{9}\piup\epsilon\sigma^3 (2R^*)^{-9}$, 
where $R^*=R/\sigma$ is a reduced particle radius. Correspondingly, the value of attractive energy is
$w = 4\piup\int^{\infty}_{2R}\,\rd r \,r^2w(r) =\frac{16}{3}\piup\epsilon\sigma^3 (2R^*)^{-3}$.
In such a case, it is conventional to define dimensionless pressure, density, and temperature by $p^*=p\sigma^3/\epsilon$, $\rho^*=\rho\sigma^3$, and  $T^*=1/ \beta\epsilon=k_{\rm B}T/\epsilon$. Thus, the equation~(\ref{eq26}) can be rewritten as
\begin{equation} \label{26-4}
p^*=T^*\rho^*+\frac{1}{2}(\rho^*)^2  \frac{16}{3}\piup (2R^*)^{-3} - \frac{1}{2}(\rho^*)^2  \frac{16}{9}\piup (2R^*)^{-9}\tan^2\Phi,
\end{equation}
and the corresponding equation for $\Phi$ is $\Phi+\frac{1}{T^*}\rho^*\frac{16}{9}\piup(2R^*)^{-9}\tan\Phi=0$. We further solve this transcendental equation  numerically to obtain the dependence of $p^*$ on $\rho^*$. 

The behavior of the curves in figure~\ref{Fig1} is in qualitative agreement with the corresponding results of reference \cite{Shi}.

\begin{figure}[h]
	\begin{center}
%\hspace{10mm}
%\begin{minipage}[htb]{0.4\linewidth}\vspace{-15mm}
\includegraphics[scale=0.7]{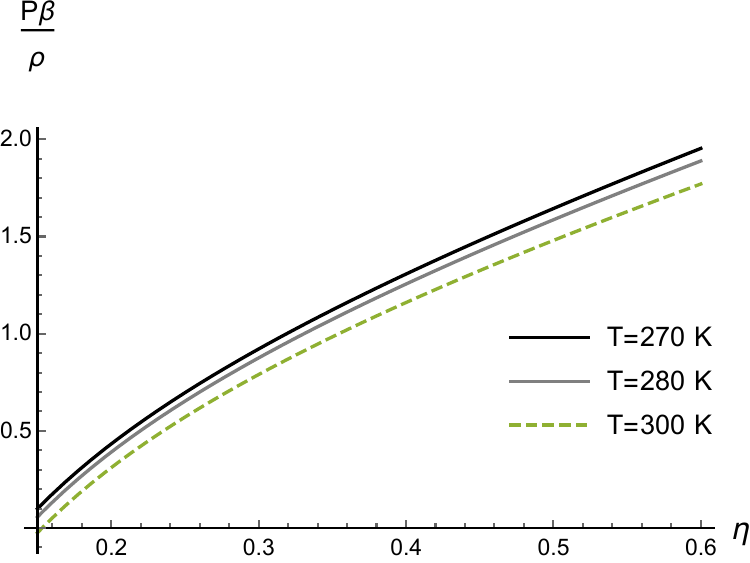} 
\caption{(Colour online) Dependence of $P\beta/{\rho}$ on $\eta$ from the equation (\ref{26-5}) for the temperature of $T=270$~K, $T=280$~K, $T=300$~K. Real units are for argon with $\sigma = 3.405$~\AA, $\epsilon/k = 119.8$~K.} \label{Fig2}
%\end{minipage}
\end{center}
\end{figure}

On the other hand, we can use the packing fraction $\eta=\frac{4}{3}\frac{N}{V}\piup R^3=
\frac{1}{6}\piup\rho\sigma^3$ since it is possible to express the atomic radius $R$ of an atom in terms of its atomic size $\sigma$. Then, the equation (\ref{eq26}) will be rewritten in the form: 
\begin{equation} \label{26-5}
\frac{P\beta}{\rho}=1+16\beta\epsilon\eta\bigg(\frac{1}{3}-\tan^2\Phi\bigg).
\end{equation}
The dependence of $\frac{P\beta}{\rho}$ on the packing fraction $\eta$ is shown in figure~\ref{Fig2}.

If  in the equation of state (\ref{eq26}), $\Phi=\frac{\piup}{2}-\delta $ and the value $\delta^2 \sim -\frac{2}{\piup} \beta u $ is small, which is possible for strong repulsive interaction on a short distance, we can rewrite the equation of state as $P= \frac{2}{\beta}\rho+ \frac{1}{2}\rho^2 w $ which fully determins the behavior of the system of particles with small repulsive interaction on a short distance and long-range attractive interaction. We can take into account the limited size of the particles in integration over the volume and the attractive interaction must increase by the order of more than three. For the systems with different long-range interactions, we use another approach. In the next section we  consider a  simple system for which a simple solution can be obtained.

\section{The systems with long-range interaction}
\label{sec-4}

In the definition of $Q_{\rm int}$ in \eqref{eq8}, the inverse operator of interaction energy is presented. In the general case of long-range interactions, such as Coulomb-like or Newtonian gravitational interaction in continuum, the limiting inverse operator should be treated as a well-known operator
\begin{equation}      \label{eq27}
U^{-1}(\mathbf{r},\mathbf{r'})=-\frac{1}{4\piup G
	m^{2}}\Delta_{\mathbf{r}}\delta(\mathbf{r}-\mathbf{r'})=L^{\psi}_{\mathbf{r}\mathbf{r'}}\delta(\mathbf{r}-\mathbf{r'}),
\end{equation}
where $\Delta_{\mathbf{r}}$ is the Laplace operator in real space. The number of realistic interactions, for which the inverse operator can be found, is limited. 
The difficulties in obtaining the inverse operator can be avoided by introducing the collective variables that correspond to the relationship between the introduced fields on the saddle-point trajectory. 

In our case of such long-range attractive and repulsive interactions between particles, we can rewrite the statistical sum of the local equilibrium distribution function in the form:
\begin{align}    \label{eq28}
Q_{l}  &=\int D\varphi D\psi \, \rd \xi \exp\bigg\{\int \frac{1}{2r_{\varphi}} \varphi(\mathbf{r}) L^{\varphi}_{\mathbf{r}\mathbf{r'}}\varphi(\mathbf{r'})+
\frac{1}{2r_{\psi}} \psi(\mathbf{r})L^{\psi}_{\mathbf{r}\mathbf{r'}}\psi(\mathbf{r'}) \nonumber \\ 
&+ \xi(\mathbf{r})\Lambda^{-3}(\mathbf{r})\exp\sqrt{\beta(\mathbf{r})}\psi(\mathbf{r})\cos \big[\sqrt{\beta(\mathbf{r})}\varphi(\mathbf{r})\big]\, \rd \mathbf{r}\bigg\},
\end{align}
where all functions $\beta(\mathbf{r})$, $\varphi(\mathbf{r})$, $\psi(\mathbf{r})$ depend on the spatial point. Here, we use the definition of interaction length as $r_{\psi}(\mathbf{r})=4\piup G m^{2}\beta(\mathbf{r})$, $r_{\varphi}(\mathbf{r})=4\piup q^{2}\beta(\mathbf{r})$. After general presentation of partition function, we can describe some real situation and determine the thermodynamic parameters for a nonequilibrium system with long-range interaction between particles.

The mean field model considering long-range interactions has been also studied. 
There is a theory which permits to quantitatively predict the particle distribution in the quasi-stationary states. Such a theory is applied to various long-range interacting systems, ranging from plasmas to self-gravitating clusters and kinetic spin models \cite{Levin}. The quantum-field-theory approach was used to give a statistical description of a system of interacting particles with arbitrary spatially inhomogeneous configurations. The formation of structures in a Coulomb-like system was analyzed and applied to the case of  dusty crystals and two-dimensional colloidal crystals \cite{Lev4}.

First of all we consider the situation when only a repulsive interaction between particles exists. The behavior of Coulomb-like system in equilibrium case was well  described  in reference \cite{Lev4}. Now, we will study only the peculiarities of the system behavior under nonequilibrium conditions. In this case, the repulsive interaction $\psi = 0$ exists. We can rewrite the statistical sum of the local equilibrium partition function in the form: 
\begin{equation}  \label{eq29} 
Q_{l}=\int D\varphi\, \rd \xi \exp   \left\{\int   \left[\frac{1}{2} \varphi(\mathbf{r})L_{\mathbf{r}\mathbf{r'}}\varphi(\mathbf{r'})+
\xi(\mathbf{r})\Lambda^{-3}(\mathbf{r})\cos \big(\sqrt{\beta(\mathbf{r})}\varphi(\mathbf{r})\big)\right]   \rd\mathbf{r}\right\}.
\end{equation}
In general case, the statistical sum $Q_{l}$ of the local equilibrium partition function is given by the expression
\begin{equation}     \label{eq30}
Q_{l} = \int D\varphi\, \rd \xi \exp\left[S(\varphi(\mathbf{r}),\xi(\mathbf{r}),\beta(\mathbf{r}))\right],
\end{equation}
with the effective nonequilibrium local thermodynamic potential $S$:
\begin{equation}     \label{eq31}     
S=\int \left\{\frac{1}{2}\varphi(\mathbf{r})L_{\mathbf{r}\mathbf{r'}}\varphi(\mathbf{r'})+
\xi(\mathbf{r})\Lambda^{-3}(\mathbf{r})\cos \big[\sqrt{\beta(\mathbf{r})}\varphi(\mathbf{r})\big]\right\}\rd\mathbf{r}.
\end{equation}
From normalization condition $\int\rho(\mathbf{r})\,\rd\mathbf{r}=N$, one can obtain the macroscopic density function $\rho(\mathbf{r})$:
\begin{equation}     \label{eq32}
\rho(\mathbf{r})\equiv \Lambda^{-3}(\mathbf{r}) \xi(\mathbf{r})\cos\big[\sqrt{\beta(\mathbf{r})}\varphi(\mathbf{r})\big].
\end{equation}
In the case of the system without interaction (for free particles $\varphi(\mathbf{r})=0$), if we write the chemical activity in terms of the chemical potential $\xi(\mathbf{r})=\exp [\mu(\mathbf{r})\beta(\mathbf{r})]$, we obtain the well-known relation $\beta(\mathbf{r})\mu(\mathbf{r})=\ln \rho(\mathbf{r})\Lambda^{3}(\mathbf{r})$ 
that generalizes the relation in the equilibrium statistical mechanics~\cite{Hua}.
The equation for energy conservation is modified now, and it has the next form:
\begin{equation}      \label{eq33}
\frac{1}{2}\int \frac{\rho(\mathbf{r})}{\beta(\mathbf{r})}\left\{3-
\sqrt{\beta(\mathbf{r})}\varphi(\mathbf{r})\tan\big[\sqrt{\beta(\mathbf{r})}\varphi(\mathbf{r})\big]\right\}\rd\mathbf{r}=E.
\end{equation}
Derivation of the energy-conservation equation with respect to the volume yields a relation for the chemical potential, i.e., 
\begin{equation}   \label{eq34}
\frac{1}{2}\frac{\rho(\mathbf{r})}{\beta(\mathbf{r})}\left\{3-
\sqrt{\beta(\mathbf{r})}\varphi(\mathbf{r}) \tan\big[\sqrt{\beta(\mathbf{r})}\varphi(\mathbf{r})\big]\right\}=\frac{\delta E}{\delta V}\frac{\delta V}{\delta N}=\mu(\mathbf{r})\rho(\mathbf{r}),
\end{equation}
hence, the chemical potential is 
\begin{equation}     \label{eq35}
\mu(\mathbf{r})\beta(\mathbf{r})=\frac{3}{2}-\frac{1}{2}\sqrt{\beta(\mathbf{r})}\varphi(\mathbf{r})\tan\big[\sqrt{\beta(\mathbf{r})}\varphi(\mathbf{r})\big].
\end{equation}
This approach also provides  the equation of state for the system within the context of  thermodynamic relation $P=\frac{1}{\beta}\frac{\delta S}{\delta V}$ 
for the case of energy conservation. The local equation of state takes the form:
\begin{equation}    \label{eq36}    
P(\mathbf{r}) \beta(\mathbf{r}) = \rho(\mathbf{r}) \left[\mu(\mathbf{r})\beta(\mathbf{r})-\frac{1}{2}\right].
\end{equation}
In the case of ideal gas we obtain a usual equation of state, because in this case there is no interaction and hence $\varphi(\mathbf{r})=0$. Thus,  we have $\mu(\mathbf{r})\beta(\mathbf{r})=\frac{3}{2}$, and the equation of state reproduces the equation of state of the ideal gas $P\beta=\rho $. The energy of the system is equal to $E=\frac{3}{2}Nk_{\rm B}T$, and it is in accord with the previous well-known results \cite{Hua}. Within the context of the definition of equation~(\ref{eq36}) we can conclude that, under the condition $\mu(\mathbf{r})\beta(\mathbf{r})< \frac{1}{2}$, the negative pressure $P(\mathbf{r}) \beta(\mathbf{r})< 0$ appears. Such situation is possible under the realistic condition $2<\sqrt{\beta(\mathbf{r})}\varphi(\mathbf{r}) \tan[\sqrt{\beta(\mathbf{r})}\varphi(\mathbf{r})]$, for constant temperature and for the total energy of the system $E<\frac{1}{2}Nk_{\rm B}T$. This condition implies that the energy of each particle is lower than the thermal energy. In this special case, the energy of the system is lower than the total thermal energy of particles. This very special condition may be associated with the very peculiar properties.

If we take into account the definition of the chemical potential, we can rewrite the density in the form:
\begin{equation}      \label{eq37}
\rho(\mathbf{r})\equiv \Lambda_{e}^{-3}(\mathbf{r}) \exp \left[\frac{1}{2}\sigma(\mathbf{r}) \tan\sigma(\mathbf{r})\right] \cos \sigma(\mathbf{r}),
\end{equation}
where the new variable $\sigma=\sqrt{\beta(\mathbf{r})}\varphi(\mathbf{r})$ is introduced and $\Lambda_{e}=\left[{\hbar^{2}\beta(\mathbf{r})e}/{2m} \right]^{\frac{1}{2}}$ is the renormalized de~Broglie wavelength.
In the average field approximation, using the local thermodynamic potential in terms of the new variable for constant temperature, we can write the local thermodynamic potential in such a form:
\begin{equation}     \label{eq38}
S=\int \left\{\frac{1}{2\beta}\sigma(\mathbf{r})L_{\mathbf{r}\mathbf{r'}}\sigma(\mathbf{r'})+
\Lambda_{e}^{-3}\exp \left[-\frac{1}{2}\sigma(\mathbf{r}) \tan\sigma(\mathbf{r})\right] \cos \sigma(\mathbf{r})\right\}\rd\mathbf{r}.
\end{equation}
Now, the equation for the field variable can be rewritten as follows:
\begin{equation}     \label{eq39}
L_{\mathbf{r}\mathbf{r'}}\sigma(\mathbf{r'})-\beta \frac{\rd V(\sigma)}{\rd\sigma}=0,
\end{equation}
where the potential energy $V=\Lambda_{e}^{-3}\exp \left[-\frac{1}{2}\sigma(\mathbf{r})\tan\sigma(\mathbf{r})\right] \cos \sigma(\mathbf{r})$
is a function of the field variable in the present form. This potential has the minimum under condition $3\sin 2\sigma=-2\sigma$. For a small value of $\sigma $ we have two different solutions: $\sigma=0$ and $\sigma^{2}=1$. For small $\sigma$ the effective potential takes a very simple form: $V(\sigma)= 1-\sigma^{2}$, and the equation for the field variable is $L_{\mathbf{r}\mathbf{r'}}\sigma(\mathbf{r'})+2\sigma(\mathbf{r'})=0$. In the case of Coulomb-like interaction, the solution for the  field is of the oscillation character. 

In the general, formula (\ref{eq31}) can describe the condition of new phase formation, the size of the bubble, and other parameters of the thermodynamic behavior of such systems. This nonequilibrium statistical description concerns only probable dilute structures of such systems, and it does not describe the metastable states and does not give any information about the time scales in the dynamic theory. In this way, however, we can solve the complicated problems of  statistical description of interacting systems. To this end, we must determine the dynamic equation for the field. In this sense we can use the Ginsburg--Landau equation to introduce the field in a standard form:
\begin{equation}   \label{eq40}
\frac{\partial \sigma(\mathbf{r},t) }{\partial t}=-\gamma \frac{\delta S}{\delta \sigma(\mathbf{r})}=-\gamma \frac{\rd V(\sigma)}{\rd \sigma},
\end{equation}
where $\gamma$ is dynamic viscous coefficient \cite{Lif,Gor,Cahn}. This evolution equation is  in fact applicable to a number of systems with nonconserved order parameter. In such a case, all necessary conditions satisfy the thermodynamic relation. We can suppose that the motivation of such a dynamics is an increase of the local thermodynamic potential. The evolution in nonequilibrium case will be guided by the local thermodynamic potential landscape and the morphological instabilities of the parameter. The dynamics of the system is dissipative. It will lead to a decrease of the local thermodynamic potential \cite{Bal}. 

\subsection{Self-gravitating system}
Interesting situations where only attractive interaction is presented are not numerous. One of the examples of such situations is the behavior of a self-gravitating system. The solution of saddle-point equation completely determines all thermodynamic parameters for  attractive field $\psi$ and describes the general behavior of self-gravitating system for both spatially homogeneous and inhomogeneous particle distributions. The above set of equations in principle solves the many-particle problem in the thermodynamic limit. It is very important to note that only this approach makes it possible to take into account the inhomogeneity of temperature distribution that may depend on the spatial distribution of the particles in the systems. In other approaches, the dependence of the temperature on a spatial point is introduced through the polytrophic dependence of temperature on particle density in the equation of state~\cite{Pad,Chav,Cha,Lyn,Sas}. In the present approach, such a dependence follows from the necessary thermodynamic condition, and can be found for various particle distributions. Now, we derive the saddle-point equation for the extreme of the local thermodynamic functional $S(\psi,\xi,\beta)$ from the statistical sum of the local equilibrium distribution function: 
\begin{equation}      \label{eq41}
Q_{l}  =  \int    D\varphi D\psi \, \rd \xi  \exp  \left\{\int \left[\frac{1}{2r_{\psi}} \psi(\mathbf{r})L^{\psi}_{\mathbf{r}\mathbf{r'}}\psi(\mathbf{r'})+
\xi(\mathbf{r})\Lambda^{-3}(\mathbf{r})\exp\left(\sqrt{\beta(\mathbf{r})}\psi(\mathbf{r})\right)\right]  \rd \mathbf{r}\right\}.
\end{equation}
The equation for the field variable ${\delta S}/{\delta \psi} = 0$ 
in the case of the absence of repulsive interaction $\varphi = 0$ yields
\begin{equation}      \label{eq42}
\frac{1}{r_{m}}\Delta
\psi(\mathbf{r})+\xi(\mathbf{r})\left[\frac{2\piup
m}{\hbar^{2}\beta(\mathbf{r})}
\right]^{\frac{3}{2}}\sqrt{\beta(\mathbf{r})}\exp\big[\sqrt{\beta(\mathbf{r})}\psi(\mathbf{r})\big]=0,
\end{equation}
where $r_{m}\equiv 4\piup G m^{2}$. The normalization condition 
can be written as follows:
\begin{equation}       \label{eq43}
\int \xi(\mathbf{r})\left[\frac{2 m}{\hbar^{2}\beta(\mathbf{r})}
\right]^{\frac{3}{2}}\exp\big[\sqrt{\beta(\mathbf{r})}\psi(\mathbf{r})\big]\rd\mathbf{r}=N,
\end{equation}
and the equation for the energy conservation in the system is given by
\begin{equation}       \label{eq44}
\frac{1}{2}\int \left[\frac{2\piup m}{\hbar^{2}\beta(\mathbf{r})}
\right]^{\frac{3}{2}}\frac{\xi(\mathbf{r})}{\beta(\mathbf{r})}\left[3-
\sqrt{\beta(\mathbf{r})}\psi(\mathbf{r})\right]\exp\big[\sqrt{\beta(\mathbf{r})}\psi(\mathbf{r})\big]\rd\mathbf{r}=E.
\end{equation}
To get more information about the behavior of a self-gravitating system, we introduce new variables. The normalization condition $\int \rho(\mathbf{r})\,\rd\mathbf{r}=N$ yields the definition for the density function, i.e., 
\begin{equation}      \label{eq45}
\rho(\mathbf{r})\equiv \left[\frac{2\piup
m}{\hbar^{2}\beta(\mathbf{r})}
\right]^{\frac{3}{2}}\xi(\mathbf{r})\exp\big[\sqrt{\beta(\mathbf{r})}\psi(\mathbf{r})\big],
\end{equation}
which reduces the equations to a simpler form. The equation for the field variable is given by
\begin{equation}      \label{eq46}         
\Delta
\psi(\mathbf{r})+r_{m}\sqrt{\beta(\mathbf{r})}\rho(\mathbf{r})=0.
\end{equation}
In the case of constant temperature and chemical activity, this equation transforms into equation for gravitational potential $\psi=\sqrt{\beta(\mathbf{r})}\psi$ in the
well-known form:
\begin{equation}     \label{eq47}
\Delta \psi(\mathbf{r})=-4\piup G m^{2}\beta \rho(\mathbf{r}).
\end{equation}
First of all we consider the equilibrium case, where all parameters are independent of the space coordinate. Thus, the equation for distribution of particles 
leads to a simple condition $\sqrt{\beta}\rho =0$, that can be realized only 
for $T\rightarrow \infty$. The particle distribution in a self-gravitating system can be homogeneous only at very high temperatures. The equation for energy conservation takes the form:
\begin{equation}       \label{eq48}
\frac{1}{2}\int \frac{\rho(\mathbf{r})}{\beta(\mathbf{r})}\left[3-
\sqrt{\beta(\mathbf{r})}\psi(\mathbf{r})\right]\rd\mathbf{r}=E.
\end{equation}
The obtained equations cannot be solved in the general case, but it is possible to analyse many cases of the behavior of a self-gravitating system under various external conditions. Hereinafter we obtain the chemical activity in terms of the chemical potential $\xi(\mathbf{r})=\exp \left[\mu(\mathbf{r})\beta(\mathbf{r})\right]$.
After differentiation of the equation for energy conservation with respect to the volume, we obtain the relation for the chemical potential: 
\begin{equation}       \label{eq49}
\frac{1}{2}\frac{\rho(\mathbf{r})}{\beta(\mathbf{r})}\left[3-
\sqrt{\beta(\mathbf{r})}\psi(\mathbf{r})\right]=\frac{\delta E}{\delta V}=\frac{\delta E}{\delta N}\frac{\delta N}{\delta V}=\mu(\mathbf{r})\rho(\mathbf{r}).
\end{equation}
The chemical potential is given by the relation:
\begin{equation}      \label{eq50}
\mu(\mathbf{r})\beta(\mathbf{r})=\frac{3}{2}-\frac{1}{2}\sqrt{\beta(\mathbf{r})}\psi(\mathbf{r}).
\end{equation}
Using the expression for density, the definition of reduced thermal de~Broglie wavelength and the gravitation length, i.e., 
\begin{equation}     \label{eq51}
\Lambda(\mathbf{r})=\left[\frac{\hbar^{2}\beta(\mathbf{r})}{2m e}
\right]^{\frac{1}{2}}, \hspace{5mm} R_{g}(\mathbf{r})=2\piup G m^{2}\beta(\mathbf{r}),
\end{equation}
we can rewrite all equations and the normalization condition in terms of density and temperature. Thus, we have:
\begin{equation}     \label{eq52}
\Delta \left\{\frac{\ln[\Lambda^{3}(\mathbf{r})\rho(\mathbf{r})]}{\sqrt{\beta(\mathbf{r})}}\right\}+
\frac{R_{g}(\mathbf{r})}{\sqrt{\beta(\mathbf{r})}}\rho(\mathbf{r})=0.
\end{equation}
Some solutions of this equation under special conditions were obtained in reference~\cite{Lev1}. An interesting case is the case when only particle density depends on the coordinate while the temperature is fixed. In this case, the equation for density takes the form:
\begin{equation}     \label{eq53}
\Delta \left[\ln \Lambda^{3}(\mathbf{r})\rho(\mathbf{r})\right]+R_{g}\rho(\mathbf{r})=0.
\end{equation}
The above equation (\ref{eq53}) can be transformed to $\Delta \left[\ln\rho(\mathbf{r})\right]+R_{g}\rho(\mathbf{r})=0$, which has an exact solution, $\rho(\mathbf{r})={2}/{R_{g}\mathbf{r}^{2}}$ but the normalization condition holds only for the case of a fixed box of the size $R={NGm^{2}}/{4k_{\rm B}T}$, fixed energy $E=Nk_{\rm B}T$. The change of chemical-potential density within the box is given by $\mu=k_{\rm B}T\left(\frac{3}{2}-{2 \Lambda^{3}}/{4k_{\rm B}TR_{g}r^{2}}\right)$. If we introduce $f(\mathbf{r})=\ln\rho(\mathbf{r})$, the present equation can be transformed to Lane--Emden equation in the form \cite{Lev,Pad,Chav,Cha}: $
\Delta f(\mathbf{r})+R_{g}\exp f(\mathbf{r})=0$ which has an exact solution for particle density only in the one-dimensional case: $\rho(\mathbf{r}) =  1/{\cosh^{2}( \mathbf{r}/ R_{g})}$. This solution has a simple form, but unfortunately it is not in a good agreement  with the results of molecular dynamic simulations \cite{Sas} and with general behavior of spatial inhomogeneous distribution of particles in self-gravitating systems. More details of such peculiarites were presented in the articles \cite{Lev,Lev1,Lev2}. 

Using the definition of the density of particles, one can obtain the local thermodynamic potential in the form:
\begin{equation}     \label{eq54}
S=\int \left\{-\rho(\mathbf{r}) \ln \left[\Lambda^{3}(\mathbf{r})\rho(\mathbf{r}) \right]- \rho(\mathbf{r})\right\}\rd\mathbf{r},
\end{equation}
from which the local equation of state can be obtained:
\begin{equation}     \label{eq55}
P(\mathbf{r}) \beta(\mathbf{r}) = -\frac{1}{\beta}\frac{\delta S}{\delta V}=\rho(\mathbf{r}) \left\{1-\ln [\Lambda^{3}(\mathbf{r})\rho(\mathbf{r})]\right\}=\rho(\mathbf{r}) \left[\mu(\mathbf{r})\beta(\mathbf{r})-\frac{1}{2}\right].
\end{equation}
From equation (\ref{eq55}) we can easily get the chemical potential $\mu(\mathbf{r})\beta(\mathbf{r})= \frac{3}{2}-\ln \left[\Lambda^{3}(\mathbf{r})\rho(\mathbf{r})\right]$ if the thermodynamic relation for conservation energy $E$ is used. In the classical case $\Lambda^{3}(\mathbf{r})\rho(\mathbf{r})\ll 1$ and $P\beta \sim \rho $ but the expression has a multiplier which logarithmically depends on the density of the particles. Only in the case where $\Lambda^{3}(\mathbf{r})\rho(\mathbf{r})= 1$, we obtain the equation of state for an ideal gas. We see a sense to talk about pressure in classical case only. If the concentration is large, and the reverse relation $1\ll \Lambda^{3}(\mathbf{r})\rho(\mathbf{r})$ takes place, the determination of pressure is not correct. There is a natural limit of our approach. We cannot describe the processes which can be realized within short distances, since in a system with large concentration the quantum effects take place. In the case of an ideal gas, we obtain usual equation of state, because in this case $\varphi(\mathbf{r})=0$ and $P\beta=\rho $ as a result of the absence of the interaction. In the case of ideal gas $\mu(\mathbf{r})\beta(\mathbf{r})=\frac{3}{2}$, and the equation of state reproduces the equilibrium relation for an ideal gas. In this case the energy of the system is $E=\frac{3}{2}Nk_{\rm B}T$, which corresponds to the previously obtained results \cite{Hua,Isi}. 

\begin{figure}[htb]
\centerline{\includegraphics[width=0.65\textwidth]{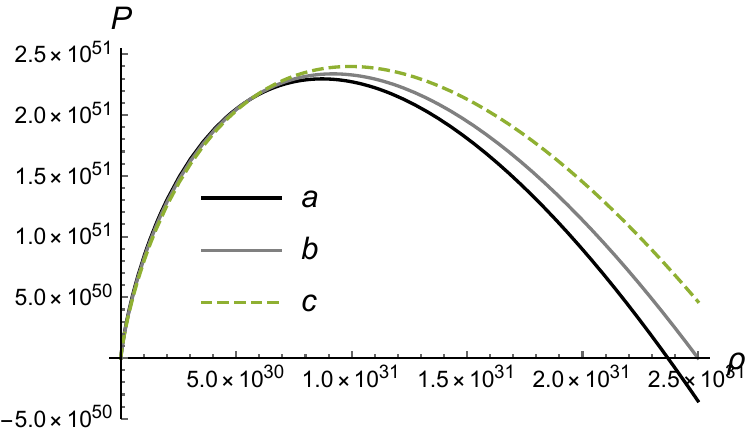}}
\caption{(Colour online) The dependence of $P$ on $\rho$ for protons ($m=1.6726\times10^{-27}$~kg) for different temperatures a) $T=275$~K, b) $T=285$~K, c) $T=300$~K.} \label{Fig3}
\end{figure}

In figure~\ref{Fig3} we show the dependence of $P$ on $\rho$ from equation (\ref{eq55}). 

Next, we consider the statistical induced dynamics of a self-gravitating system where the possible spatial distribution of particles is taken into account. To this end, we must determine the dynamic equation for the field variable or density of the particles. In this sense we can use the Ginsburg--Landau equation for density: 
\begin{equation}    
\frac{\partial \rho(\mathbf{r},t) }{\partial t}=-\nabla^{2} \gamma \frac{\delta S}{\delta \rho(\mathbf{r})},  \label{eq56}
\end{equation}
where $\gamma$ is the dynamic gravitational viscous coefficient \cite{Chan}. In a self-gravitating system we do not have  any other viscous process without dynamic influence of gravitational action of all the system on the local spatial point. This evolution equation is, in fact, applicable to a number of systems with non-conserved order parameter.
If we take into account the expression for the local thermodynamic potential in terms of density, we can write the dynamic equation for density as follows:
\begin{equation}      \label{eq57}
\frac{\partial \rho(\mathbf{r},t) }{\partial t}=-\nabla^{2} \gamma \ln \left[\Lambda^{3}(\mathbf{r})\rho(\mathbf{r},t)\right].
\end{equation}
The evolution of patterns in a nonequilibrium case is guided by the local thermodynamic potential landscape and the morphological instabilities of the parameter. The dynamics of the system is dissipative. It will result in a decrease of the local thermodynamic potential of the patterns with time. To describe this behavior, we can use the simple gradient descent
dynamics defined by the chemical potential. It can be considered as generalization of Cahn equation for a non-uniform system with an arbitrary concentration gradient which becomes Cahn nonlinear diffusion equation \cite{Cahn}:
\begin{equation}      \label{eq58}
\frac{\partial \rho(\mathbf{r},t) }{\partial t}=\nabla \gamma \nabla \beta \mu(\mathbf{r})=-\nabla^{2} \gamma \ln [\Lambda^{3}(\mathbf{r})\rho(\mathbf{r},t)].
\end{equation}
This equation (\ref{eq58}) is fully equivalent to the previously obtained dynamic equation in Ginsburg--Landau approach.
If we take the solution of this dynamic equation in the form $\rho(\mathbf{r},t)= \exp (\gamma r_{m} t)\rho(\mathbf{r}) $, the solution of the problem reduces to the solution of the previous equation \mbox{$\Delta \left[\ln\Lambda^{3}\rho(\mathbf{r})\right] + R_{g}\rho(\mathbf{r})=0$} which can be transformed to $ \Delta \left[\ln\rho(\mathbf{r})\right]+R_{g}\rho(\mathbf{r})=0$ for constant temperature. This means that the inhomogeneous spatial solution will exist for a permanently increasing density of particles. If we determine the possible structure in distribution of density, these patterns will be conserved during all the time of a decrease of density. In this approach, the probable behavior of a self-gravitating system can be predicted for any external condition. 

\section{Conclusions}
\label{sec-Concl}

Statistical description of a spatially inhomogeneous quasi-equilibrium system is proposed using the ideas of the method of nonequilibrium statistical operator and the Hubbard--Stratonovich transformation. It is important to note that the statistical mechanics of the systems in local equilibrium state is based on the laws of conservation not of the average values of dynamic variables, but of dynamic variables. To determine the thermodynamic function of quasi-equilibrium systems, one should use the representation of the corresponding statistical ensembles that take into account the local equilibrium states of the systems.The local equilibrium ensemble should be defined according to the distribution function or statistical operator of the system. Using an approach based on the statistical operator, we can obtain the necessary relations which are correct for both cases of homogeneous and inhomogeneous particle distribution, which is also confirmed in figures~\ref{Fig1}, \ref{Fig2} and \ref{Fig3}, respectively.

The saddle point method was used to find the asymptotic value of the statistical sum for an increasing number of particles. The dominant contribution is given by the states that satisfy the extremum condition for the functional. The saddle point equation represents a thermodynamic relation that can be written as an equation for the field variables. The spatially inhomogeneous solution of these equations corresponds to the distribution of interacting particles. It is very important to note that only in this approach one can take into account the inhomogeneous temperature distribution and chemical potential, which may depend on the spatial distribution of particles in the system.

The local equilibrium partition function with spatially inhomogeneous functions is introduced to describe systems with both attractive and repulsive interaction.
 
The case is interesting where only the particle density depends on the coordinate, and the temperature is fixed. We obtain an equation for the density, a local equation of state. We take into account the statistics of the induced dynamics of the self-gravitating system, where the possible spatial distribution of particles is taken into account. The dynamics of the system is dissipative. This will lead to a decrease in the local thermodynamic potential of the structures with time. To describe this behavior, we use simple gradient descent dynamics defined by the chemical potential. This can be considered as a generalization of the Cahn equation for a heterogeneous system with an arbitrary concentration gradient, which becomes the Cahn nonlinear diffusion equation. This means that the heterogeneous spatial solution will exist for a constant increase in the particle density. If we define a possible structure in the density distribution, these patterns will persist throughout the entire time of density decrease. In this approach, the probable behavior of a self-gravitating system can be predicted under any external conditions.

Spatially inhomogeneous distribution is not equilibrium in the thermodynamic limit and thermodynamic equilibrium is reached during the relaxation time. The Ginzburg-Landau equation describes the relaxation of an arbitrary parameter to equilibrium using the minimum of the locally equilibrium distribution. As for the Cahn equation, it takes into account the inhomogeneous distribution of the chemical potential, which is the driving force of diffusion.

\appendix
\section{Appendix}
\renewcommand*\theequation{A.\arabic{equation}}

For the system of particles with different character of interaction, the Hamiltonian is
\begin{equation}    \nonumber 
H=\sum_{i}\frac{\mathbf{p}_{i}^{2}}{2m}-\frac{1}{2}\sum_{i,j} W(\mathbf{r_{i}},\mathbf{r_{j}})+\frac{1}{2}\sum_{i,j}
U(\mathbf{r_{i}},\mathbf{r_{j}}).
\end{equation}
Taking into account that $n(\mathbf{r})=\sum_{i}\delta(\mathbf{r}-\mathbf{r_{i}})$ and the fact that the masses of particles can  differ from each other, the energy density of the system can be written as follows:
\begin{equation} \nonumber  
H(\mathbf{r})=\sum_i \frac{\mathbf{p}_i^{2}}{2{m}}{\delta(\mathbf{r}-\mathbf{r}_{i})}-\frac{1}{2}\int
W(\mathbf{r},\mathbf{r'})n(\mathbf{r})n(\mathbf{r'})\,\rd \mathbf{r'}+\frac{1}{2}\int
U(\mathbf{r},\mathbf{r'})n(\mathbf{r})n(\mathbf{r'})\,\rd \mathbf{r'}.
\end{equation}
For the system having a different character of interaction after simple mathematical manipulation using Hubbard--Stratonovich transformation, we can write the statistical sum of the local equilibrium distribution function in terms of the additional fields:
\begin{align}    
Q_{l} & = \int D\Gamma \bigg\{-\int
\bigg[\beta(\mathbf{r})\sum_i\frac{{\mathbf{p}_i}^{2}}{2{m}}{\delta(\mathbf{r}-\mathbf{r}_{i})}-\eta(\mathbf{r})n(\mathbf{r})
\bigg]\rd\mathbf{r}\,  \nonumber \\ 
&+\frac{1}{2}\iint
\beta(\mathbf{r})\left[ W(\mathbf{r},\mathbf{r'})- U(\mathbf{r},\mathbf{r'})\right]n(\mathbf{r})n(\mathbf{r'})\,\rd\mathbf{r}\,\rd \mathbf{r'}\bigg\}.
\end{align}
The integration over the phase space gives $D\Gamma=\left[{1}/{\left( 2\piup \hbar\right)^{3}}\right]\prod\limits_{i}\rd {\mathbf{r}}_{i}\, \rd{\mathbf{p}}_{i}$. Since we consider the system with nonequilibrium particle distribution, in our approximation this means, that every point of the space has its own temperature and chemical potential. Namely, temperature and chemical potential depend on the space point (coordinate). However, every volume $V$ contains a sufficient number of particles \cite{Zub}. That permits to perform first integration over the momentum separately. 

In order to perform a formal integration in the second part of our paper, 
additional field variables can be introduced. We use here the theory of Gaussian 
integrals \cite{Str,Kle}:
\begin{align}     \label{eq59}
&\exp \bigg[ -\frac{\nu ^2}{2}\iint
\beta(\mathbf{r})\omega(\mathbf{r},\mathbf{r'})n(\mathbf{r})n(\mathbf{r'})\,\rd
\mathbf{r}\, \rd \mathbf{r'}\bigg]     \nonumber  \\ 
& =\int D \sigma \exp \bigg[
-\frac{\nu ^2}{2}\iint
\omega(\mathbf{r})^{-1}\sigma(\mathbf{r})\sigma(\mathbf{r'})\,\rd
\mathbf{r}\,\rd \mathbf{r'}-\nu \int\sqrt{\beta(\mathbf{r})}\sigma(\mathbf{r})n(\mathbf{r})\,\rd
\mathbf{r} \bigg],
\end{align}
where $D\sigma={\prod\limits_{s}\rd\sigma _{s} }/{\sqrt{\det
		2\piup \beta \omega(\mathbf{r},\mathbf{r'})}}$ and
$\omega^{-1}(\mathbf{r},\mathbf{r'})$ is the inverse operator which satisfies the condition
$\int_{r'}\omega^{-1}(\mathbf{r},\mathbf{r'})\omega(\mathbf{r'},\mathbf{r''})=\delta
(\mathbf{r}-\mathbf{r''})$. This means that the interaction energy is presented by the Green 
function. For this operator, the value $\nu ^2=\pm 1$ depends on the sign of the interaction 
or the potential energy. After such manipulation the field variable 
$\sigma(\mathbf{r}) $ contains the same information as original distribution 
function, i.e., all information about possible spatial states of the systems. 

Now the statistical sum of the local equilibrium distribution function can be rewritten as follows:
\begin{align}    \label{eq60}
Q_{l} & = \int D\Gamma \int D\varphi \int D\psi \exp\bigg\{-\int
\beta(\mathbf{r})\sum_i\frac{{\mathbf{p}_i}^{2}}{2{m}}{\delta(\mathbf{r}-\mathbf{r}_{i})}\,\rd\mathbf{r} +\int\big[\eta(\mathbf{r})+\sqrt{\beta(\mathbf{r})}\psi(\mathbf{r}) \nonumber  \\ &+\ri\sqrt{\beta(\mathbf{r})}\varphi(\mathbf{r})\big]n(\mathbf{r})\,\rd\mathbf{r}\bigg\}Q_{\rm int},
\end{align}
where the expression for $Q_{\rm int}$, which corresponds to the interaction, is
\begin{equation}        \label{eq61}
Q_{\rm int}   =   \exp   \left[\frac{1}{2}\iint W(\mathbf{r},\mathbf{r'})^{-1}\psi(\mathbf{r})\psi(\mathbf{r'})\,\rd\mathbf{r}\,\rd \mathbf{r'}-   \frac{1}{2}\iint
U(\mathbf{r},\mathbf{r'})^{-1}\varphi(\mathbf{r})\varphi(\mathbf{r'})\,\rd\mathbf{r}\,\rd \mathbf{r'}\right].
\end{equation}
In this general functional, integration can be performed on the phase space. 
Using the definition of density, we can rewrite the statistical sum of the local equilibrium distribution function as follows:
\begin{align}     \label{eq62}
Q_{l} & = \int D\varphi \int D\psi \int \frac{1}{\left( 2\piup \hbar\right)
	^{3}N!}\prod\limits_{i}\,\rd r_{i} \,\rd p_{i}\,\xi(\mathbf{r_{i}})   \nonumber \\ 
& \times
{\exp \bigg\{-
\bigg[\beta(\mathbf{r_{i}})\frac{\mathbf{p}_{i}^{2}}{2m}-\sqrt{\beta(\mathbf{r_{i}})}\psi(\mathbf{r_{i}})-\ri\sqrt{\beta(\mathbf{r_{i}})}\varphi(\mathbf{r_{i}})\bigg]\bigg\}Q_{\rm int}},
\end{align}
where the new variable $\xi(\mathbf{r})\equiv \exp \eta(\mathbf{r})$ is introduced  
which can be interpreted as chemical activity. Now, we can perform integration over the momentum. 
The real part of the statistical sum has the form:
\begin{align}    \label{eq63}
Q_{l} & =\int D\varphi \int D\psi Q_{\rm int} \frac{1}{N!}    \nonumber \\ & \times  \prod\limits_{i}\int \rd r_{i} \,\xi(\mathbf{r_{i}})\left[\frac{2\piup {m}}{\hbar^{3}\beta(\mathbf{r_{i}})}
\right]^{\frac{3}{2}} \exp
\big[\sqrt{\beta(\mathbf{r_{i}})}\psi(\mathbf{r_{i}})\big]\cos\big[\sqrt{\beta(\mathbf{r_{i}})}\varphi(\mathbf{r_{i}})\big], 
\end{align}
then,
\begin{align}      \label{eq64}
Q_{l} &=\int D\varphi \int D\psi Q_{\rm int} \sum_{N}\frac{1}{N!}      \nonumber \\ &  \times  \left\{{\int \rd\mathbf{r} \,\xi(\mathbf{r})\left[\frac{2\piup {m}}{\hbar^{3}\beta(\mathbf{r})}
	\right]^{\frac{3}{2}}\exp \big[\sqrt{\beta(\mathbf{r})}\psi(\mathbf{r})\big]\cos \big[\sqrt{\beta(\mathbf{r})}\varphi(\mathbf{r})\big]}\right\}^{N}.
\end{align}
After that, the statistical sum of the local equilibrium distribution function takes a simple form:
\begin{align}   \label{eq65}
Q_{l} & =\int D\varphi \int D\psi Q_{\rm int}   \nonumber \\ & \times \exp\bigg\{\int \bigg[
\xi(\mathbf{r})\left(\frac{2\piup {m}}{\hbar^{3}\beta(\mathbf{r})}
\right)^{\frac{3}{2}}\exp\left(\sqrt{\beta(\mathbf{r})}\psi(\mathbf{r})\right)\cos \left(\sqrt{\beta(\mathbf{r})}\varphi(\mathbf{r})\right)\bigg]\rd\mathbf{r}\bigg\}.
\end{align}

 %\end{document}

\ukrainianpart

\title{Термодинамічне співвідношення для систем з неоднорідним розподілом частинок}
\author{А. П. Ребеш, Б. І. Лев, А. Г. Загородній}
\address{
Інститут теоретичної фізики ім. М.М. Боголюбова Національної академії наук України, вул.~Метрологічна~14б, 03143 Київ, Україна
}

\makeukrtitle

\begin{abstract}
\tolerance=3000
Для системи з неоднорідним розподілом частинок ми отримали термодинамічне співвідношення, яке залежить від координати. У нашому підході для отримання такого співвідношення ми використовуємо локально рівноважну функцію розподілу. Перш за все, ми визначили термодинамічне співвідношення для системи з однорідним розподілом частинок. Передбачено можливі особливості поведінки систем з різним характером взаємодії в нерівноважному випадку. За допомогою методу сідлової точки ми знайшли домінуючі внески в статистичну суму та отримали всі термодинамічні параметри систем з різним характером взаємодії. Поява сідлового стану в усіх системах взаємодіючих частинок при різних температурах і розподілах частинок мають однакову фізичну природу, тому їх можна описати однаковим чином. Ми розглядаємо системи з притяганням та відштовхуванням, а також самогравітуючі системи.
\keywords локально рівноважна функція розподілу, рівняння стану, самогравітуюча система, далекосяжна взаємодія

\end{abstract}
\lastpage
\end{document}